\documentclass[12pt]{article}
\usepackage{epsfig}
\usepackage{amsmath}
\usepackage{color}

\newcommand{\be}{\begin{equation}}
\newcommand{\ee}{\end{equation}}

\newcommand{\bea}{\begin{eqnarray}}
\newcommand{\eea}{\end{eqnarray}}
\newcommand{\beq}{\begin{equation}}
\newcommand{\eeq}{\end{equation}}
\newcommand{\nn}{\nonumber}

\def\la{\mathrel{\mathpalette\fun <}}
\def\ga{\mathrel{\mathpalette\fun >}}
\def\fun#1#2{\lower3.6pt\vbox{\baselineskip0pt\lineskip.9pt
\ialign{$\mathsurround=0pt#1\hfil##\hfil$\crcr#2\crcr\sim\crcr}}}

\begin{document}

\title{  Hadron collisions at ultrahigh energies:
black disk or resonant disk modes?
}
\author{
V.V. Anisovich$^+$,
V.A.  Nikonov$^{+ \diamondsuit}$
and J. Nyiri$^*$
}


\maketitle

\begin{center}
{\it
$^+$National Research Centre "Kurchatov Institute",
Petersburg Nuclear Physics Institute, Gatchina, 188300, Russia}

{\it $^\diamondsuit$
Helmholtz-Institut f\"ur Strahlen- und Kernphysik,
Universit\"at Bonn, Germany}

{\it $^*$Institute for Particle and Nuclear Physics, Wigner RCP,
Budapest 1121, Hungary}

\end{center}

\begin{abstract}

The analysis of current ultrahigh energy data for hadronic total cross
sections and diffractive scattering cross sections points to a steady
growth of the optical density with energy
for elastic scattering amplitudes in the impact parameter space, $b$.
At LHC energy the profile function of the $pp$-scattering
amplitude, $T(b)$, reaches the black disk limit at small $b$.
Two scenarios are possible at larger energies, $\sqrt{s}\ga 100$ TeV.
First, the profile function gets frozen in the black disk limit,
$T(b)\simeq 1$
while the radius of the black disk $R_{black\;disk}$
is increasing with $\sqrt s$, providing $\sigma_{tot}\sim \ln^2s$,
$\sigma_{el}\sim \ln^2s$,
$\sigma_{inel}\sim \ln^2s$.
In another scenario the profile function continues to
grow at $\sqrt{s}\ga 100$ TeV approaching the maximal value,
$T(b)\simeq 2$, that means the resonant disk mode.
We discuss features of the resonant disk mode
when the disk radius, $R_{resonant\;disk}\,$, increases
providing the growth of the total and elastic cross sections
$\sigma_{tot}\sim \ln^2s$,
$\sigma_{el}\sim \ln^2s$, but a more slow increase of inelastic
cross section,
$\sigma_{inel}\sim \ln s$.
\end{abstract}


\section{Introduction}

The data \cite{Latino:2013ued,auger} definitely confirm the previous
observations \cite{pre}, namely, that the total cross sections
increase steadily with energy ($\sigma_{tot}\sim \ln^n s$
as $1\la n\la 2$); the steady growth is observed for
$\sigma_{el}$ and $\sigma_{inel}$, while the ratio $ReA_{el}/ImA_{el}$
is small and probably decreases slowly.

Already the first indications of the cross sections growth
\cite{serp} gave start to corresponding models with the supercritical
pomeron \cite{super1,super2}. The concept of the power growth
of cross sections
($\sigma_{tot}\sim s^\Delta$ with $\Delta\simeq 0.08$)
became widely accepted in the 1980s
\cite{donn,kaid} and is discussed till now \cite{DL}
(let us note that exceeding of the Froissart bound \cite{Froi} does
not violate necessarily the general constraints \cite{azimov}).

It was shown in \cite{Gaisser,Block,Fletcher} that the power-type
growth of scattering amplitudes is dumped to $\ln^2 s$-type within the
$s$-channel unitarization.
The black disk picture with the $\ln^2 s$-growth of the
$\sigma_{tot}$ and $\sigma_{el}$ at ultrahigh energies was
suggested in the Dakhno-Nikonov model \cite{DN}.
The model can be considered as a realization of the Good-Walker
eikonal approach \cite{GW} for a continuous set of channels. Presently,
the black disk mode for hadron collisions at ultrahigh energies is
discussed extensively, see, for example,
\cite{1110.1479,1111.4984,1201.6298,1202.2016,1208.4086,ann1,Alkin:2014rfa,Troshin:2007fq,Giordano:2013iga}.

The black disk mode is usually discussed in terms of the optical
density for elastic scattering amplitude.
For the asymptotic regime such a presention was
carried out in \cite{ann2,annn}: the cross sections
$\sigma_{tot}(pp)$, $\sigma_{el}(pp)$, $\sigma_{inel}(pp)$ demonstrate
a maximal growth, $\sim\ln^2 s$, while diffractive dissociation cross
sections $\sigma_{D}(pp)$, $\sigma_{DD}(pp)$ give
a slower growth, $\sim\ln s$.

For the calculation of screening corrections in inelastic diffractive
processes at ultrahigh energies \cite{amn} the $K$-matrix technique
is more preferable. The K-matrix function $-iK(b)$ in the preLHC region
increases with energy being mainly concentrated at $b<1$ fm. The black
disk regime for the K-matrix function means its "freezing",
$-iK(b)\to 1$, in the disk area.
If  the growth of the $-iK(b)$
continues with increasing energy, the interaction
area turns into a resonant disk. In this case asymptotically
$\sigma_{tot}(pp)\sim\ln^2 s$, $\sigma_{el}(pp)\sim\ln^2 s$ with
$[\sigma_{el}(pp)/\sigma_{tot}(pp)]_{s\to\infty}\to 1$;
the resonant disk area is surrounded by a black border band
that provides
$\sigma_{inel}(pp)\sim\ln s$, $\sigma_{D}(pp)\sim\ln s$,
$\sigma_{DD}(pp)\sim\ln s$.

In the present paper we perform a comparative analysis of predictions for
ultrahigh energy diffractive processes in the framework of these two scenarios.
It is definitely seen that the data at $\sqrt{s}\sim 10$ TeV are
not sensitive to the versions of the disk: the initial stages are similar in both modes. Distinctions are seen at $\sqrt{s}\sim 10^3-10^4$ TeV.
Apparently, the study and interpretation of the
cosmic ray data at such energies  are the problems on the agenda.

\begin{figure}[ht]
\centerline{\epsfig{file=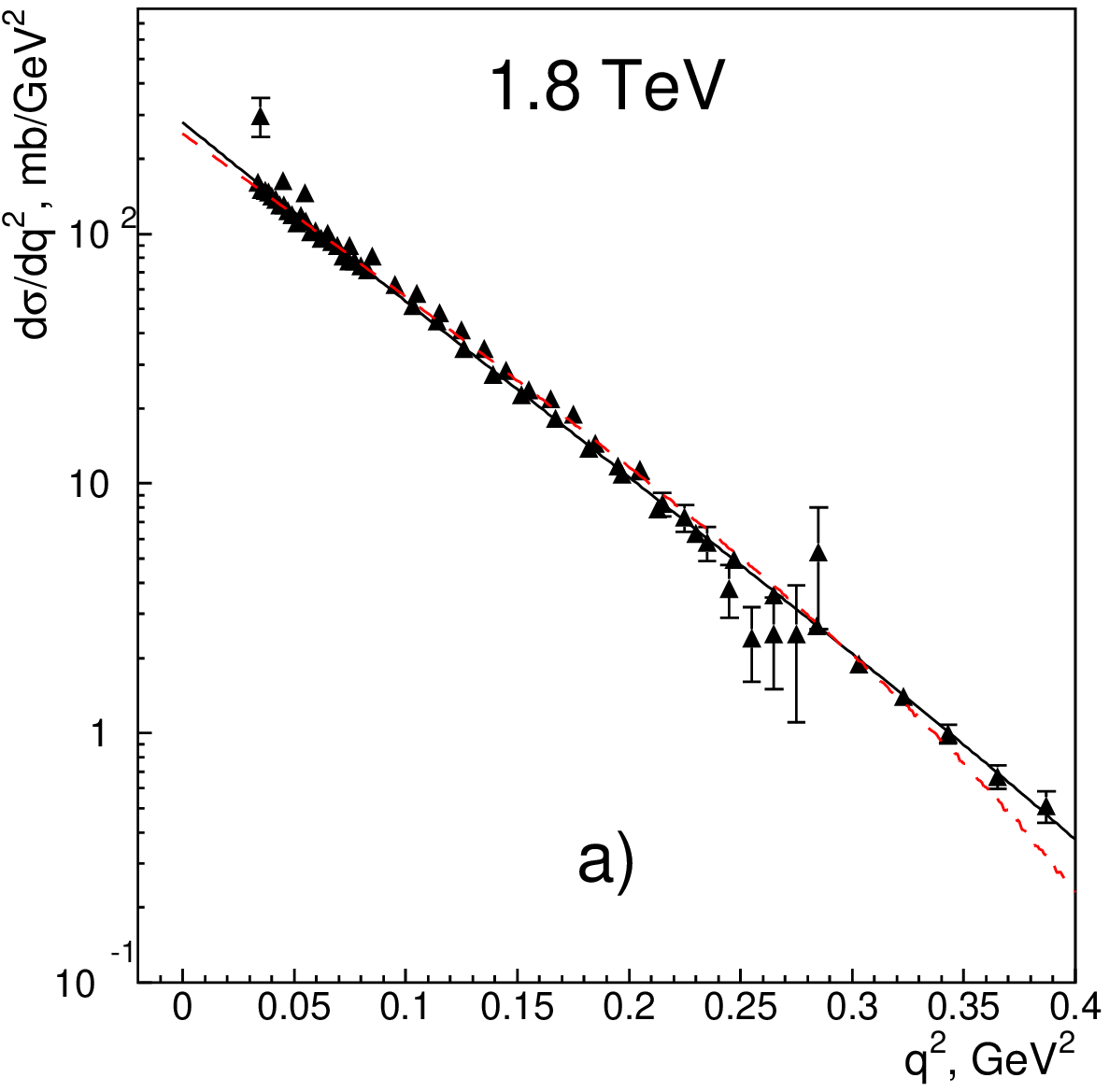,width=8cm}
            \epsfig{file=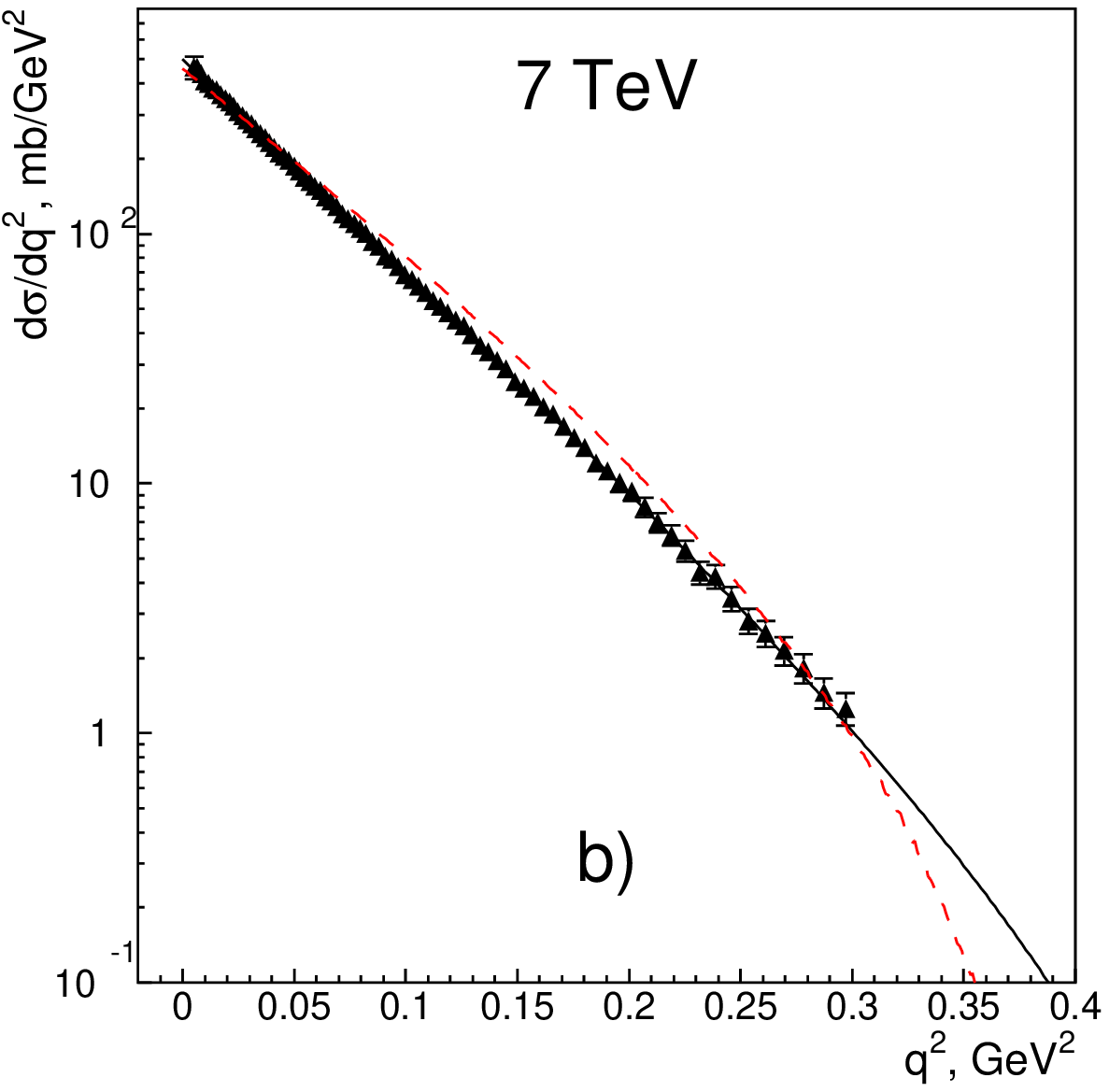,width=8cm}}
\caption{ a,b) Differential cross sections
$d\sigma_{el}/d{\bf q}^2_\perp$ at $\sqrt s = 1.8, 7.0$ TeV
and their description within the black disk mode (red dashed lines)
and the resonant disk mode (solid lines).
}
\label{rd1}
\end{figure}

\begin{figure}[ht]
\centerline{\epsfig{file=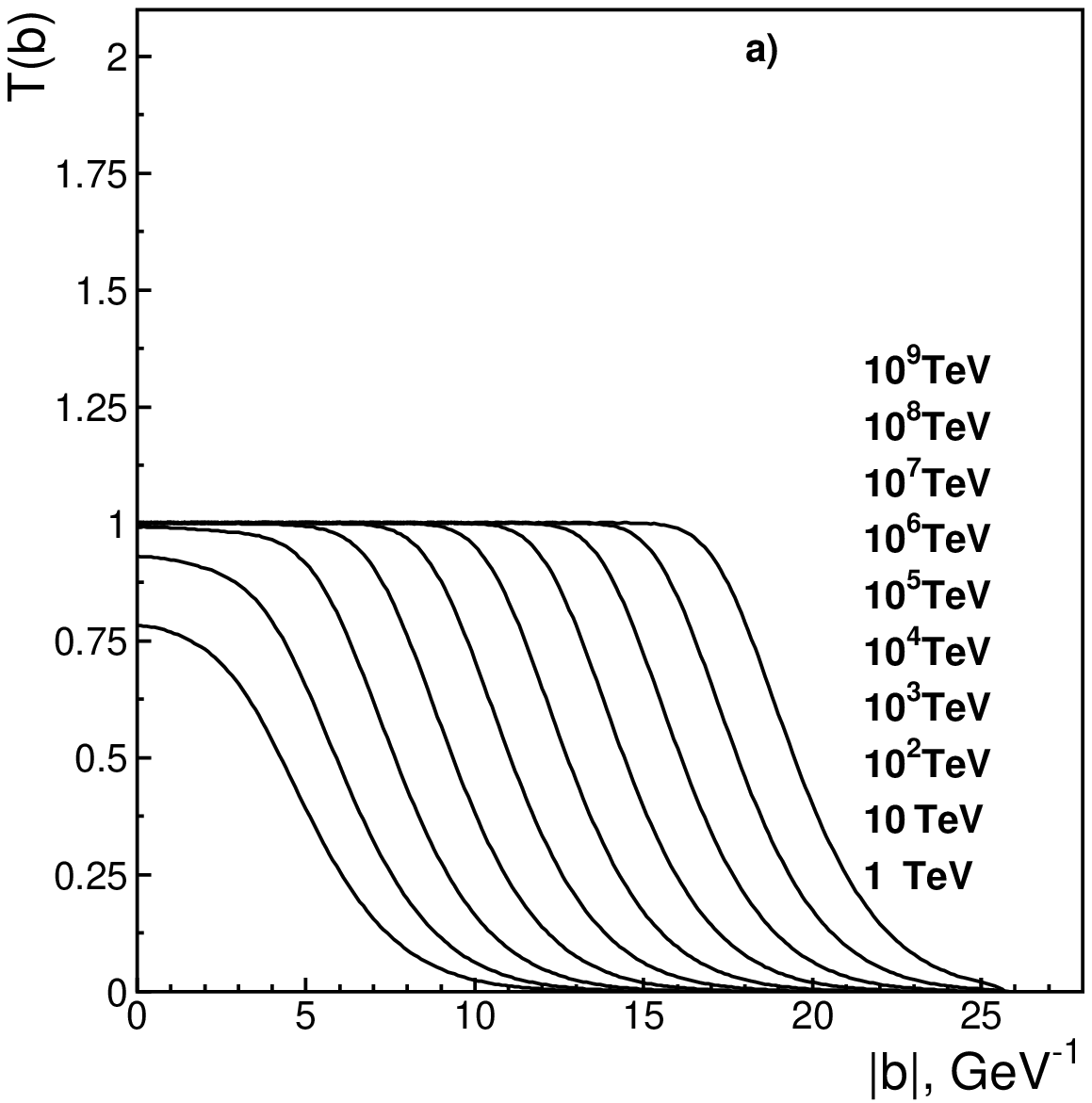,width=8cm}
            \epsfig{file=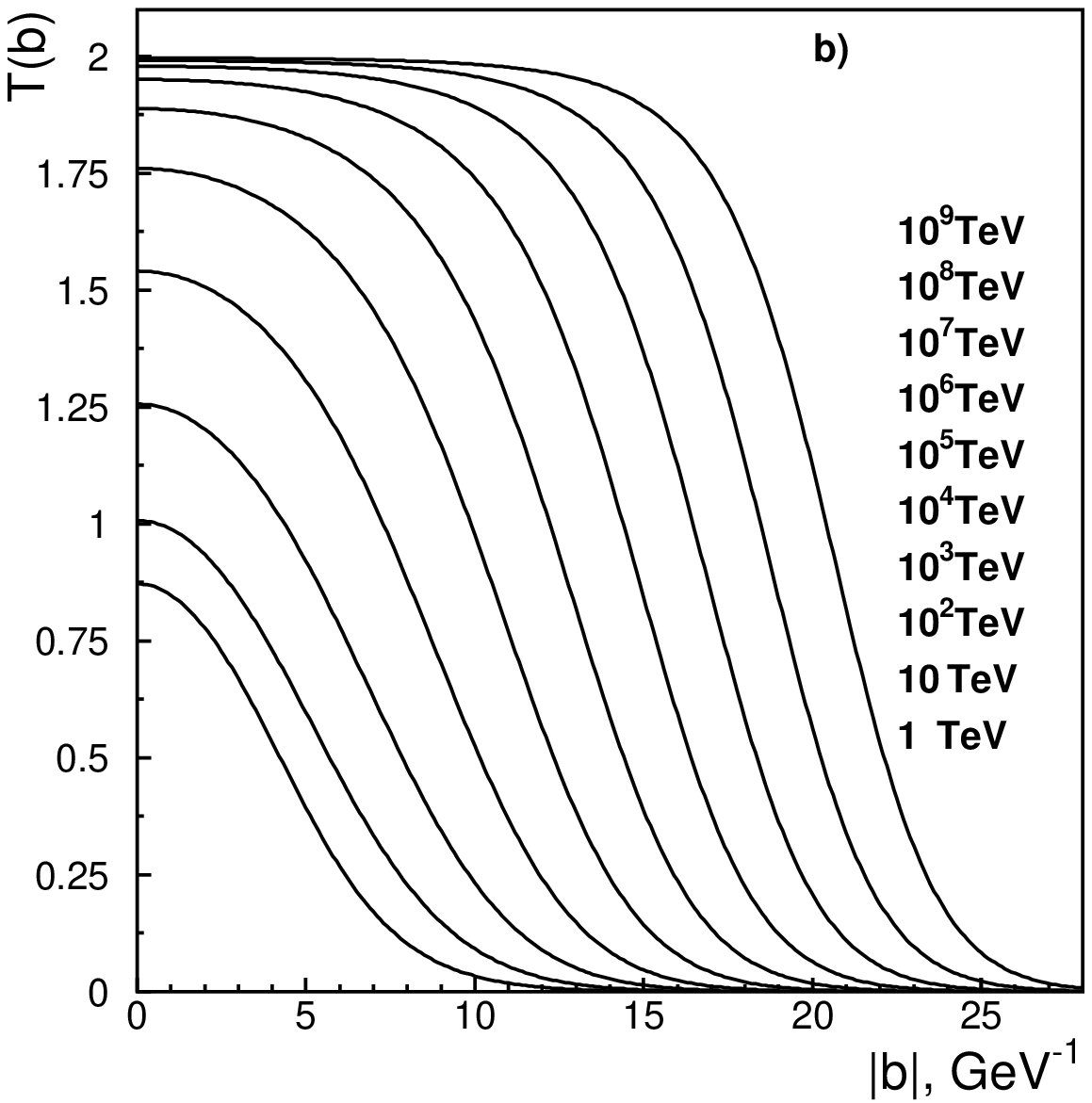,width=8cm}}
\caption {
 a)  Profile functions, $T(b)$,
 at  $\sqrt{s}=1,\,10,\,10^2,... 10^{9}$ TeV
  for the black disk regime  ($T(b)\to 1$)
 and b) resonant disk regime ($T(b)\to 2$).
 At $\sqrt{s}=1-10$ TeV the profile functions in both modes
 are nearly the same.
\label{rd2}}
\end{figure}

\section{Scattering amplitude in the impact parameter space \\
and the K-matrix representation for ultrahigh energy}

In the impact parameter space the profile function $T(b)$
is determined at high energies as:
\bea \label{1}
&&
\sigma_{tot}=2\int d^2b \; T(b)\,,
\\
&&
4\pi\frac{d\sigma_{el}}{d{\bf q}^2_\perp}=
|A_{el}({\bf q}^2_\perp)|^2,\quad
A_{el}({\bf q}^2_\perp)=i\int d^2b e^{i{\bf b}{\bf q}_\perp} T(b)\,,
\nn\\
&&
T(b)
=1-\eta(b)\, e^{2i\delta(b)}=1-e^{-\frac12\chi(b)}=
\frac{-2iK(b)}{1-iK(b)},
\nn
\eea
here $A_{el}({\bf q}^2_\perp)$ is the elastic scattering amplitude.
The profile function can be
presented either in the standard form using
the inelasticity parameter $\eta(b)$ and the phase shift
$\delta(b)$ or in terms of the optical density $\chi(b)$
and the K-matrix function $K(b)$.
The $K$-matrix approach is based on the separation of
the elastic rescatterings in the intermediate states:
the function $K(b)$ includes only
the multiparticle states thus being complex valued.
The small value of the $ReA_{el}/ImA_{el}$ tells that $K(b)$ is
dominantly imaginary.

\begin{figure}
\centerline{
            \epsfig{file=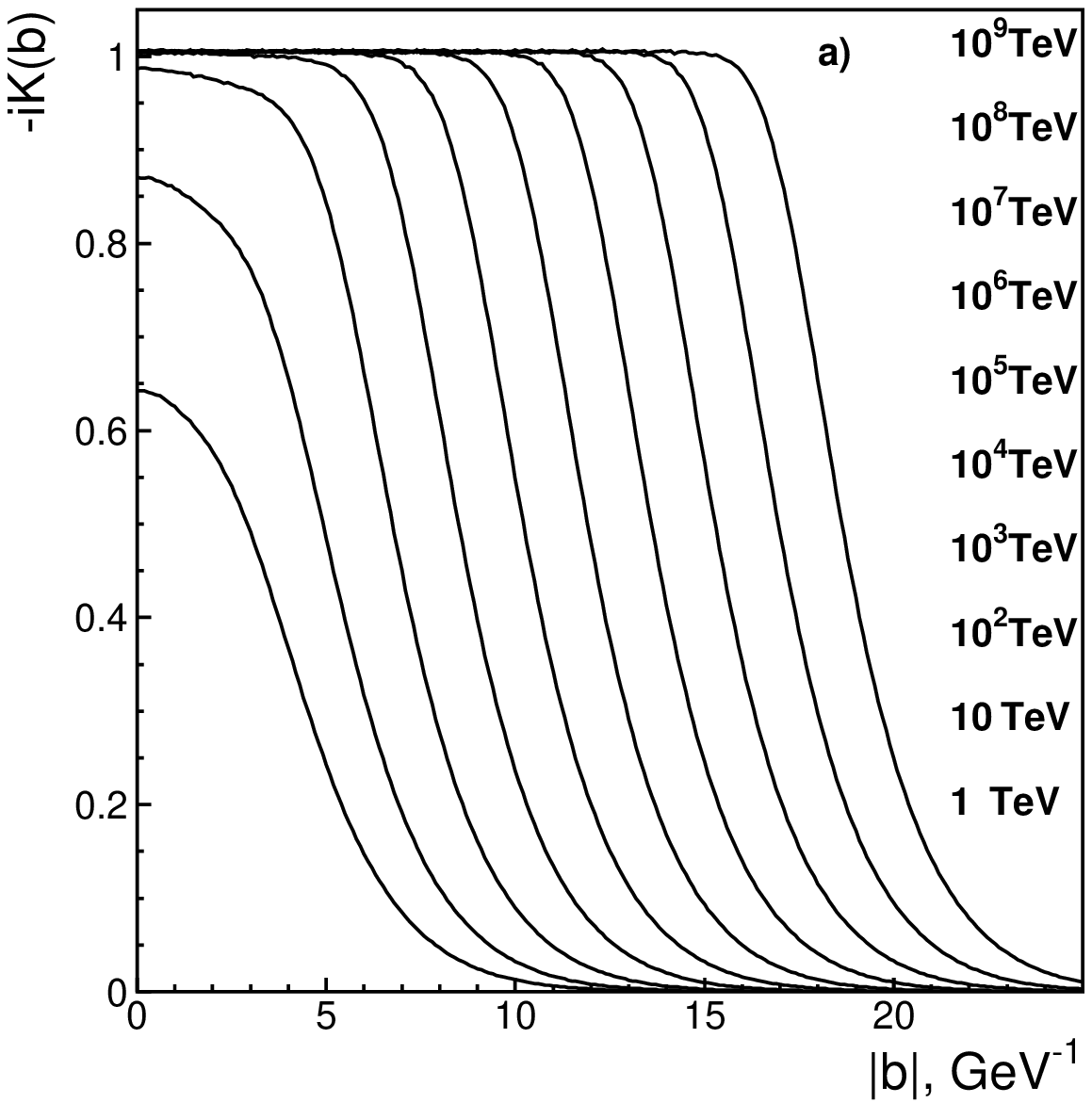,width=8cm}
            \epsfig{file=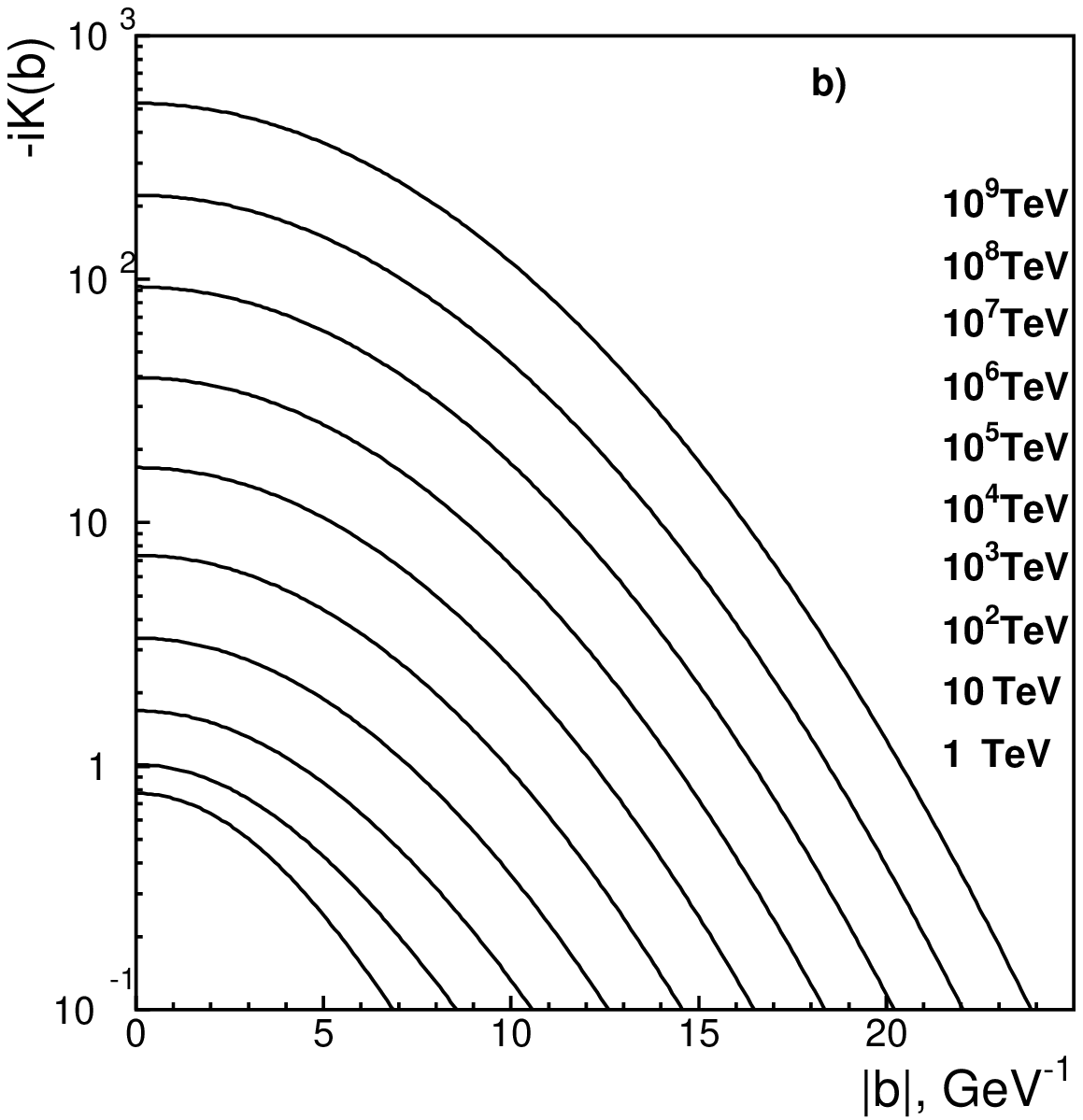,width=8cm}}
\caption{
The K-matrix functions, $-iK(b)$, for
a) the black disk mode ($[-iK(b)]_{\xi\to \infty}\to 1$ at
$b< R_0\xi$ )  and b) the
resonant disk  mode ($[-iK(b)]_{\xi\to \infty}\to \infty$ at
$b< R_0\xi$).
\label{rd3}}
\end{figure}

\begin{figure}[ht]
\centerline{\epsfig{file=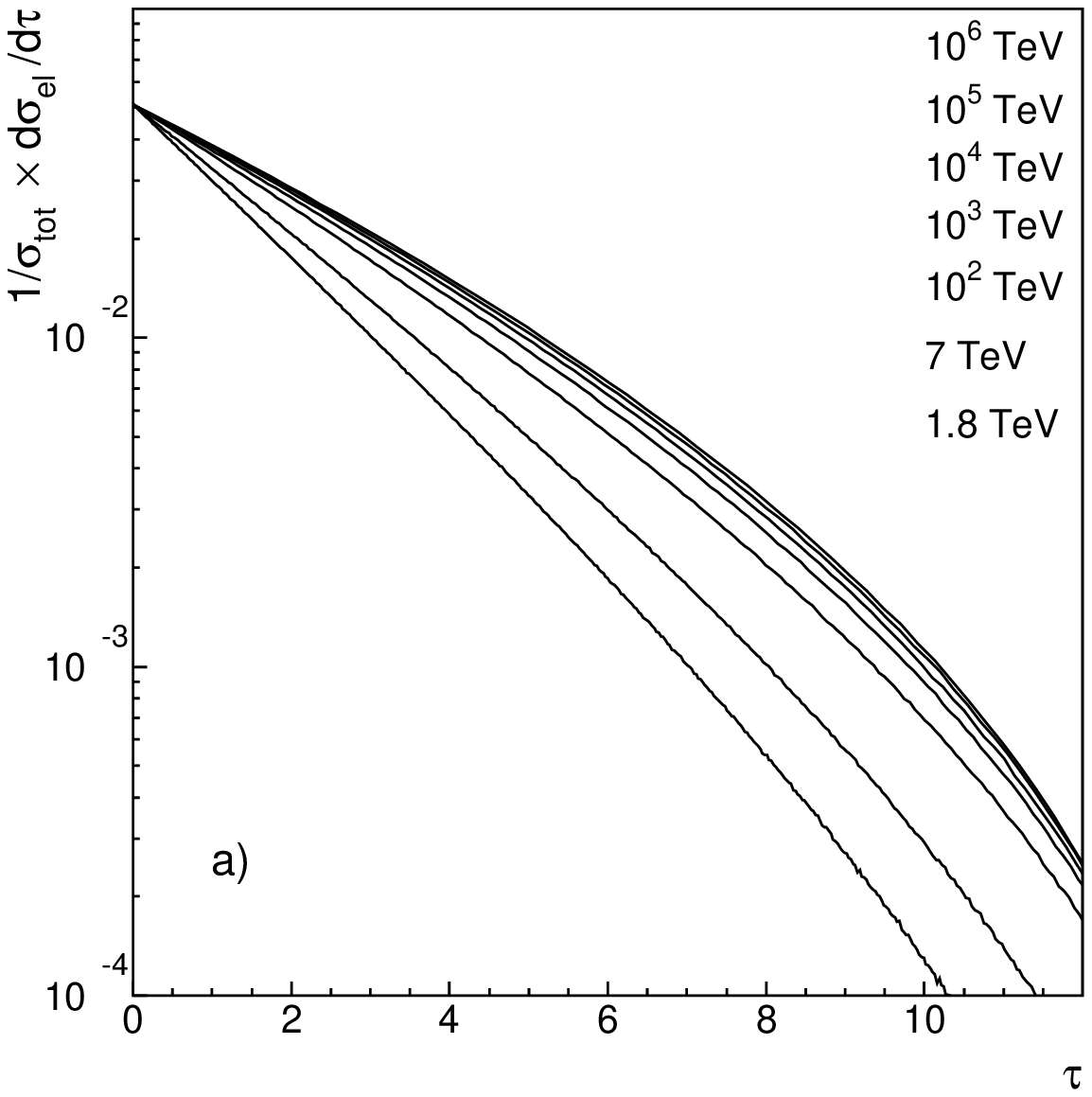,width=8cm}
            \epsfig{file=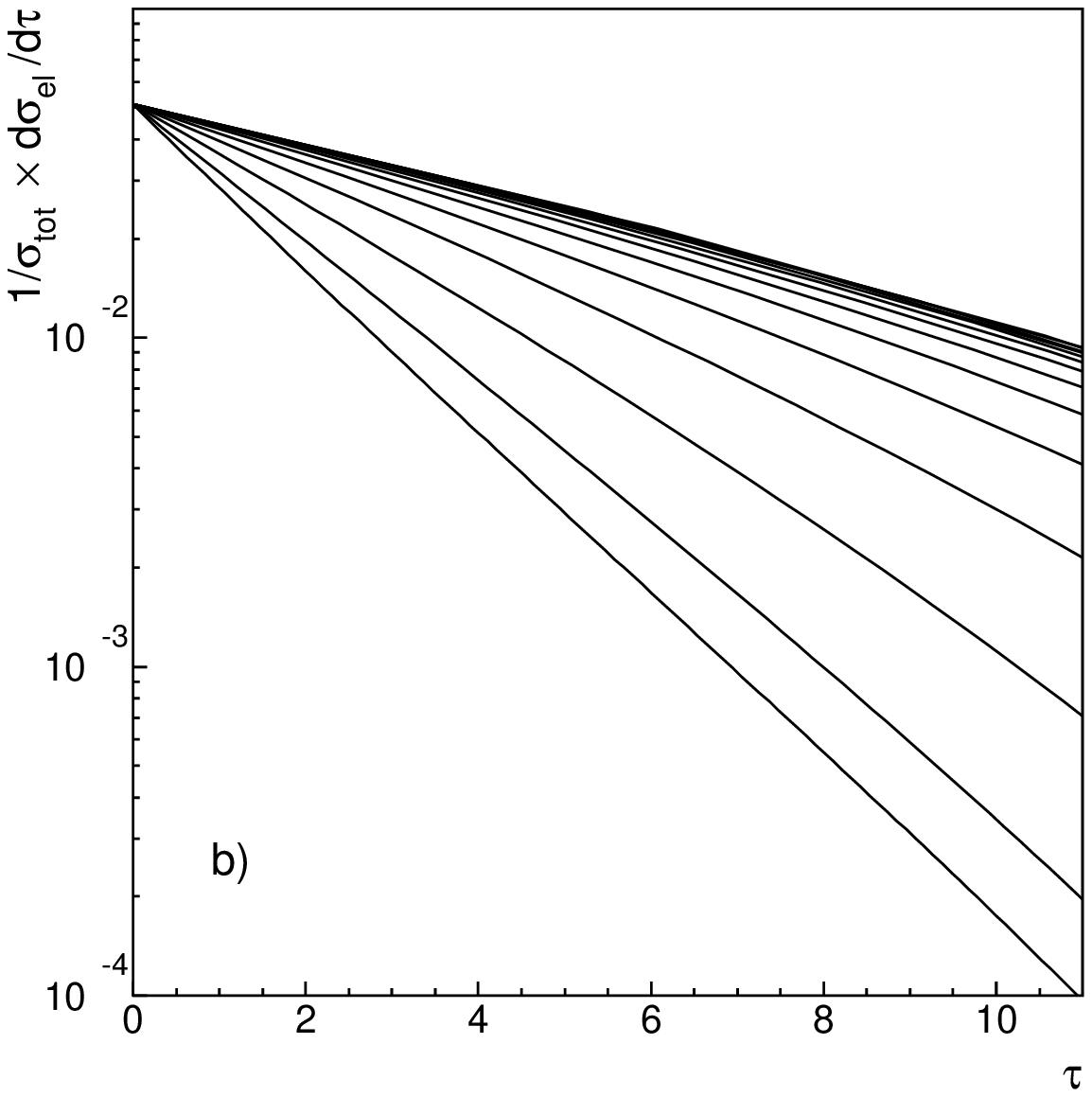,width=8cm}}
\caption{The black disk (a) and resonant disk (b) modes: the
 $\tau$-representation
($\tau=\sigma_{tot}{\bf q}^2$ ) for differential cross
section, $\frac{1}{\sigma_{tot}}\frac{d\sigma_{el}}{d\tau}$.
The differential cross sections are similar at $\sqrt{s}=1.8,\,7$
TeV; distinctions are seen at $\sqrt{s}\geq 10^3$ TeV.}
\label{rd4}
\end{figure}

\subsubsection{Black disk limit in terms of the Dakhno-Nikonov model}

The Dakhno-Nikonov model \cite{DN} demonstrates us a
representative example of application of the optical density
technique for the consideration of $pp^\pm$ collisions
at ultrahigh energies when $\ln s>>1$. In the model
the black disk is formed by the low density pomeron cloud and
rescatterings are described within the eikonal approach.
The scattering amplitude $AB\to AB$ reads:
\be \label{4}
A_{AB\to AB}({\bf q}^2)=i\int d^2b
e^{i{\bf q}{\bf b}}\int dr'\varphi^2_A(r')dr''
\varphi^2_B(r'')
\left [1-\exp{\Big(-\frac{1}{2}
\chi_{AB}(r',r'',{\bf b})}\Big)\right],
\ee
where $dr\varphi^2_{A}(r)$, $dr\varphi_{B}^2(r)$ are the quark
densities of the  colliding hadrons in the impact parameter space.
Proton and pion quark densities can be determined using the
corresponding form factors. The optical density
$\chi_{AB}(r',r'',{\bf b})$ depends on parameters of
the $t$-channel interaction.

The behavior of amplitudes at ultrahigh energies is
determined by leading complex-j singularities, in the
Dakhno-Nikonov model that are leading and next-to-leading pomerons
 with trajectories
$\alpha({\bf q}^2)\simeq 1+\Delta-\alpha'{\bf q}^2$.
The fit of refs. \cite{ann1,ann2} gives:
\be
\begin{tabular}{l|l|l}
  parameters       & leading pole  & next-to-leading    \\
\hline
$\Delta$                    & 0.27          &  0      \\
$\alpha'_P$ [(GeV)$^{-2}$] &  0.13           & 0.25    \\
\end{tabular}
\ee
In terms of the K-matrix approach
the black disk mode means the assumed freezing of the $-iK(b)$ in the interaction area:
 \bea \label{23-17}
 &&
\Big[-i K(b)\Big]_{\xi\to\infty}\to 1
 \qquad {\rm at} \; b<R_0 \, \xi\,,
 \\
&&
 \Big[-iK(b)\Big]_{\xi\to\infty}\to 0
\qquad
 {\rm at} \; b>R_0 \, \xi \, ,\nn
 \\
 \xi=\ln\frac{s}{s_R},&&\quad s_R\simeq 6.4\cdot 10^3\;{\rm GeV}^2,
 {\rm with}\;
 R_{0}\simeq 2\sqrt{\alpha'\Delta}\simeq 0.08 \; {\rm fm}.\nn
\eea
The growth of the radius of the black disk is slow: the
small value of $R_0$ is caused by the large mass of glueballs
\cite{AIP-conf,ijmp} and the effective mass of gluons
\cite{parisi,field}.

The black disk mode results in
\bea
&&
\sigma_{tot}\simeq 2\pi(R_0\xi)^2,
\\
&&
\sigma_{el}\simeq \pi(R_0\xi)^2, \quad
\sigma_{inel}\simeq \pi(R_0\xi)^2.
\nn
\eea
For the black disk radius the corrections of the order of $\ln \xi$
exist $R_{black\; disk}\simeq R_0\xi+\varrho\ln\xi$ but
they become apparent in the Dakhno-Nikonov model at energies
of the order of the Plank mass, $\sqrt{s}\sim 10^{17}$ TeV.

\begin{figure}[ht]
\centerline{
            \epsfig{file=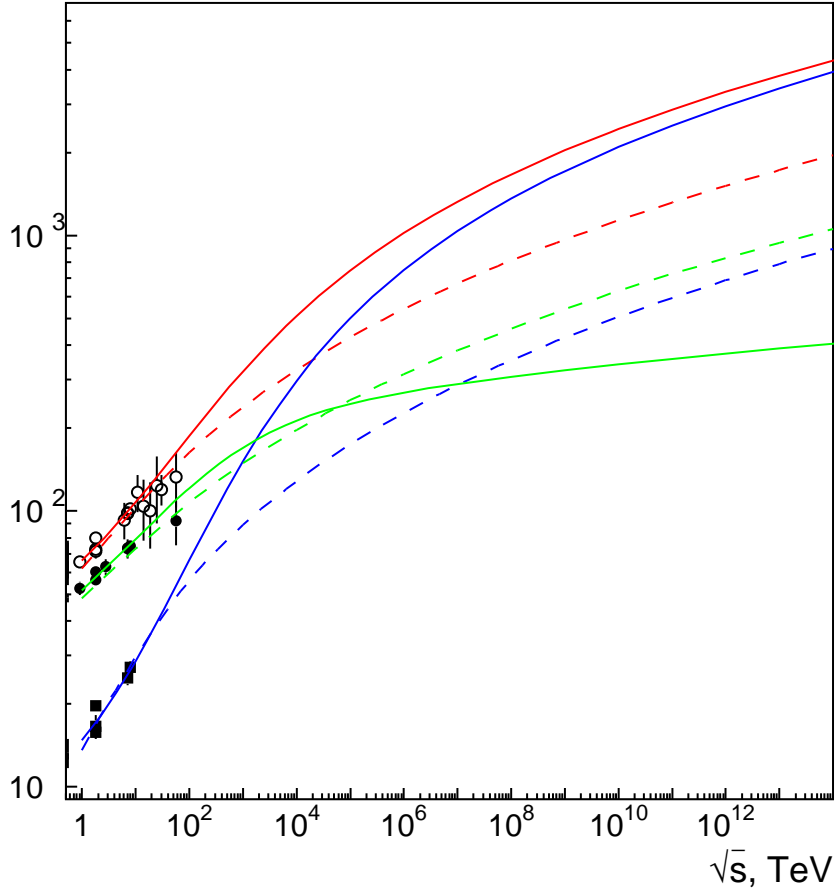,width=12cm}}
\caption{
Total, elastic and inelastic cross sections for
resonant disk (solid lines) and black disk (dashed lines) modes:
red $\sigma_{tot}$, blue $\sigma_{el}$,
green $\sigma_{inel}$.
}
\label{rd5}
\end{figure}

\subsubsection{Resonant disk and K-matrix function  }

From the data it follows that both $T(b)$ and $-i K(b)$
are increasing with energy, being less
than unity. If the eikonal mechanism does not quench the
growth, both characteristics cross the black disk limit getting
 $T(b)>1$, $-i K(b)>1$. If $-i K(b)\to \infty$ at $\ln s\to \infty$,
which corresponds to a growth caused by the supercritical
pomeron ($\Delta>0$), the
diffractive scattering process gets to the resonant disk mode.

For following the resonant disk switch-on we use the two-pomeron
model with parameters providing the description of data at 1.8 TeV and 7 TeV, namely:
\bea
\label{rd5a}
&&
-i K(b)=\int \frac{d^2q}{(2\pi)^2}
\exp{\Big(-i{\bf q}{\bf b}\Big)}
\sum g^2 s^\Delta
e^{-(a+\alpha\xi){\bf q}^2)}
\\
&&
= \sum
\frac{g^2}{4\pi(a+\alpha'\xi)}
\exp{\Big[\Delta\xi-\frac{{\bf b}^2}{4(a+\alpha'\xi)}\Big]}\,,
\qquad \xi=\ln \frac {s}{s_0}.
\nn
\eea
The following parameters are found
for the leading and the next-to-leading pomerons:
\be
\begin{tabular}{l|l|l}
  parameters       & leading pole  & next-to-leading    \\
\hline
$\Delta$                    &  0.20          &  0      \\
$\alpha'_P$ [GeV$^{-2}$] &  0.18           &  0.14   \\
$a$   [GeV$^{-2}$]     &  6.67           & 2.22   \\
$g^2$  [ mb ]        &  1.74           &  28.6  \\
$s_0$  [GeV$^{2}$]   &  1               &   1   \\
\end{tabular}
\ee
The description of the differential cross sections
$d\sigma_{el}/d{\bf q}^2_\perp$ at $\sqrt s = 1.8, 7.0$ TeV
in resonant disk mode is demonstrated in Fig. \ref{rd1}.
%
The resonant interaction regime occurs at
$b<2\sqrt{\alpha'\Delta}\xi=R_0\xi$, in this region $T(b)\to 2$.
In terms of the inelasticity parameter and the phase shift
it corresponds to $\eta\to 1$ and $\delta\to \pi/2$.
Cross sections at $\xi\to\infty$ obey
$\sigma_{tot}\simeq 4\pi R^2_0 \xi^2$, $\sigma_{el}/\sigma_{tot}\to 1$
and $\sigma_{inel}\simeq 2\pi R_0 \xi$.

\subsubsection{Comparative survey
of the resonant disk and black disk modes}

At the energy $\sqrt s\sim 10$ TeV the black cloud fills out the
proper hadron domain, the region $\leq 1$ fm, and that happens in
both modes.
It is demonstrated in Figs. \ref{rd2},\ref{rd3}: the profile functions $T(b)$ coincide practically
in both modes as well as the K-functions $-iK(b)$.
 Correspondingly, the differential cross sections in
$\tau$-representation differ a little, mainly at $\tau\sim 10$,
Fig. \ref{rd4}. The energy behavior of
$\sigma_{tot}$, $\sigma_{el}$ and $\sigma_{inel}$
coincide also at $\sqrt s\sim 1- 100$ TeV in both modes,
Fig. \ref{rd5}.

Differences appear at $\sqrt s\sim 1000$ TeV: $T(b)\simeq 1.5$
at $b\la 0.5$ fm and the black zone has shifted to
$b\simeq 1.0-1.5$ fm, Fig. \ref{rd3}b. With further energy increase
the radius of the black band increases as
$2\sqrt{\Delta\alpha'}\xi\equiv R_{rd}\xi$. The rate of growth
in both modes is determined by the leading singularity and the fit
of the data in the region $\sqrt s\sim 1-10 $ TeV gives approximately the same values of $\Delta$ and $\alpha'$ for both  cases thus providing $R_0\simeq R_{rd}$.

%
%
\section{Conclusion}

The interaction of soft gluons determines the physics of hadrons.
The effective gluons are massive and their mass is of the order
of 1 GeV that is seen directly in radiative decays of heavy
quarkonia \cite{parisi,field}, $\psi\to \gamma+ hadrons$ and
$\Upsilon\to \gamma+ hadrons$.
The effective gluon mass is determinative both for low energy physics, making possible to introduce the notion of the
constituent quark, and for high energy physics, dictating the
rate of the growth of the  interaction radius. High energy physics is the physics of large logarithms, $\ln s/s_0>>1$, and
the value $\sqrt{s_0}\sim m_{effective\, gluon}$ corresponds
to a start of the asymptotic regime at $\sqrt{s} \sim 1$ TeV. However, the initial increments of the measured characteristics
such as  $\sigma_{tot}$ ,
$\sigma_{el}$ and $\sigma_{inel}$ are visually similar, and  therefore their behavior in this region does not distinguish
between different versions. A real discrimination of modes can
appear when cross section data are discussed at much larger energies,
$\sqrt{s} \sim 10^3-10^4$ TeV.

Cosmic ray data probably can provide information to fix asymptotic
mode. Another way is to study the diffractive inelastic processes which differ strongly for different modes \cite{amn}.

\subsubsection*{Acknowledgment}

We thank  M.G. Ryskin
and A.V. Sarantsev for useful discussions and comments.
 The work was supported by grants RFBR-13-02-00425 and
 RSGSS-4801.2012.2.


\begin{thebibliography}{99}

\bibitem{Latino:2013ued}
  G.~Latino [on behalf of TOTEM Collaboration],
  EPJ Web Conf.\  {\bf 49}, 02005 (2013)
  [arXiv:1302.2098 [hep-ex]].

\bibitem{auger} Pierre Auger Collaboration (P. Abreu {\it et al}.),
Phys. Rev. Lett. {\bf 109}, 062002 (2012).


\bibitem{pre}
UA4 Collaboration, Phys. Lett. {\bf B147 }, 385 (1984);\\
UA4/2 Collaboration, Phys. Lett. {\bf B316}, 448 (1993);\\
UA1 Collaboration, Phys. Lett. {\bf B128}, 336 (1982);\\
E710 Collaboration, Phys. Lett. {\bf B247}, 127 (1990);\\
CDF Collaboration, Phys. Rev. {\bf D50}, 5518 (1994).

\bibitem{serp}Y.P. Gorin {\it et al}.
Yad. Phys. {\bf 14}, 998 (1971).

\bibitem{super1}
 A. Capella and J. Kaplan, Phys. Lett. {\bf B52}, 448 (1974).

\bibitem{super2} P.E. Volkovitsky, M.A. Lapidus,
V.I. Lisin, K.A. Ter-Martirosyan,
Yad. Phys. {\bf 24}, 1237 (1976).

\bibitem{donn}
A. Donnachie and P.V. Landshoff, Nucl. Phys. {\bf B231}, 189 (1984).

\bibitem{kaid}A.B. Kaidalov and K.A. Ter-Martirosyan,
Sov. J. Nucl. Phys. {\bf 39}, 979 (1984).

\bibitem{DL}A. Donnachie and P.V. Landshoff,
arXiv:11122485, (2011) [hep-ph].

\bibitem{Froi} M. Froissart, Phys. Rev. {\bf 123}, 1053 (1961).

\bibitem{azimov} Y.I. Azimov,
  Phys. Rev. {\bf D84}, 056012 (2011);
arXiv:1208.4304(2012)  [hep-ph].

\bibitem{Gaisser}  T.K. Gaisser and T. Stanev, Phys. Lett.,
{\bf B219}, 375, 1989.

\bibitem{Block} M. Block, F. Halzen and B. Margolis, Phys.
Lett., {\bf B252},
 481, 1990.
\bibitem{Fletcher}R.S. Fletcher, Phys. Rev. {\bf D46}, 187, 1992.

\bibitem{DN} L.G. Dakhno and V.A. Nikonov,
Eur. Phys. J. {\bf A8}, 209 (1999).


\bibitem{GW}M.L. Good, W.D. Walker, Phys. Rev. {\bf 120}, 1857
(1960).

%

\bibitem{1110.1479}F. Halzen, K. Igi, M. Ishida and C.S. Kim,
Phys. Rev. {\bf D85}, 074020 (2012);
arXiv:1110.1479V2(2012) [hep-ph].

\bibitem{1111.4984}V. Uzhinsky and A. Galoyan,
 arXiv:1111.4984v5(2012) [hep-ph].

\bibitem{1201.6298} M.G. Ryskin, A.D. Martin and V.A. Khoze,
Eur. Phys. J. {\bf C72}, 1937(2012);
  arXiv:1201.6298v2(2012)   [hep-ph].

\bibitem{1202.2016}I.M. Dremin, V.A. Nechitailo,
Phys. Rev. {\bf D85}, 074009 (2012);
  arXiv:1202.2016 (2012)  [hep-ph].

\bibitem{1208.4086}M.M. Block and F. Halzen,
Phys. Rev. {\bf D86}, 0501504 (2013);
  arXiv:1208.4086v1 (2012)  [hep-ph].

\bibitem{ann1}V.V. Anisovich, K.V. Nikonov, and V.A. Nikonov,
Phys. Rev. {\bf D88}, 014039 (2013);
 [arXiv:1306.1735  [hep-ph]].

\bibitem{Alkin:2014rfa}
  A.~Alkin, E.~Martynov, O.~Kovalenko and S.~M.~Troshin,
  Phys.\ Rev.\ {\bf D89}, 091501 (2014)
  [arXiv:1403.8036 [hep-ph]].

\bibitem{Troshin:2007fq}
  S.~M.~Troshin and N.~E.~Tyurin,
  Int.\ J.\ Mod.\ Phys.\ {\bf A22}, 4437 (2007)
  [hep-ph/0701241].

\bibitem{Giordano:2013iga}
  M.~Giordano and E.~Meggiolaro,
  JHEP {\bf 1403}, 002 (2014)
  [arXiv:1311.3133 [hep-ph]].

\bibitem{ann2} V.V. Anisovich, V.A. Nikonov, and J. Nyiri,
Phys. Rev. {\bf D88}, 014039 (2013);
 [arXiv:1310.2839 (hep-ph)].

\bibitem{annn} V.V. Anisovich, K.V. Nikonov,  V.A. Nikonov
 and J. Nyiri,
Int. J. Mod. Phys. {\bf A29}, 1450096 (2014);
 arXiv:1404.1904 (hep-ph).




\bibitem{amn} V.V. Anisovich, M.A. Matveev, and V.A. Nikonov,
{\it Hadron diffractive production at ultrahigh energies},
  arXiv:1407.4588 (hep-ph).

















\bibitem{AIP-conf}V.V. Anisovich, AIP Conf. Proc. {\bf 619}, 197 (2002),
{\bf 717}, 441 (2004);
Phys.Usp. {\bf 47}, 45 (2004), [UFN {\bf 47}, 49 (2004)].

\bibitem{ijmp} V.V. Anisovich, M.A. Matveev,
  J. Nyiri, A.V. Sarantsev,
Int. J. Mod. Phys. {\bf A20}, 6327 (2005).

\bibitem{parisi}G. Parisi and R. Petronzio,
Phys. Lett. {\bf 94}, 51 (1980).

\bibitem{field}M. Consoli and J.H. Field,
Phys. Rev. {\bf D49}, 1293 (1994).

\end{thebibliography}
\end{document}